\documentclass[sn-apa, iicol]{sn-jnl}

\jyear{2023}%

\theoremstyle{thmstyleone}%
\theoremstyle{thmstyletwo}%

\theoremstyle{thmstylethree}%
\usepackage{soul}
\usepackage{rotating}
\usepackage{xcolor}
\usepackage{amsmath,amssymb,amsfonts}
\usepackage{array}
\usepackage{subfigure}
\usepackage{graphicx}
\usepackage{textcomp}
\usepackage{multirow}
\usepackage{multicol}

\usepackage{booktabs}

\begin{document}

\title[Article Title]{MultiFusionNet: Multilayer Multimodal Fusion of Deep Neural Networks for Chest X‑Ray Image Classification}






\author[1]{\fnm{Saurabh} \sur{Agarwal}}\email{saurabha@iiitm.ac.in}

\author[2]{\fnm{K. V.} \sur{Arya}}\email{kvarya@iiitm.ac.in}

\author*[3]{\fnm{Yogesh Kumar} \sur{Meena}} \email{yk.meena@iitgn.ac.in}

\affil[1]{\orgdiv{Department of Computer Science and Engineering}, \orgname{ABV-Indian Institute  of Information Technology \& Management, Gwalior,} \country{India}}

\affil[2]{\orgdiv{Department of Computer Science and Engineering}, \orgname{ABV-Indian Institute  of Information Technology \& Management, Gwalior,} \country{India}}

\affil[3]{\orgdiv{Department of Computer Science and Engineering}, \orgname{Indian Institute of Technology Gandhinagar}, \orgaddress{\country{India}}}


\abstract{Chest X-ray imaging is a critical diagnostic tool for identifying pulmonary diseases. However, manual interpretation of these images is time-consuming and error-prone. Automated systems utilizing convolutional neural networks (CNNs) have shown promise in improving the accuracy and efficiency of chest X-ray image classification. While previous work has mainly focused on using feature maps from the final convolution layer, there is a need to explore the benefits of leveraging additional layers for improved disease classification. Extracting robust features from limited medical image datasets remains a critical challenge. In this paper, we propose a novel deep learning-based multilayer multimodal fusion model that emphasizes extracting features from different layers and fusing them. Our disease detection model considers the discriminatory information captured by each layer. Furthermore, we propose the fusion of different-sized feature maps (FDSFM) module to effectively merge feature maps from diverse layers. The proposed model achieves a significantly higher accuracy of 97.21\% and 99.60\% for both three-class and two-class classifications, respectively. The proposed multilayer multimodal fusion model, along with the FDSFM module, holds promise for accurate disease classification and can also be extended to other disease classifications in chest X-ray images.}




\maketitle

\section{Introduction}
\label{sec:introduction}
 
Respiratory infections and pneumonia share significant similarities in their causes. Specifically, pneumonia accounts for 14\% of all deaths among children under five years old, resulting in the loss of 740,180 young lives in 2019~\cite{PneumoniaDashboard}. Given the resemblance of respiratory infections, early-stage diagnosis becomes challenging. Accurate diagnosis plays a crucial role in ensuring appropriate treatment. The real-time reverse-transcriptase-polymerase chain reaction (rRT-PCR) test is currently employed for COVID-19 detection. However, this method can be time-consuming, taking hours to days, depending on available resources. Thus, developing a detection method that offers immediate and accurate results at an affordable cost is essential. Fundamental diagnostic techniques such as Chest X-Ray (CXR) and Computed Tomography (CT) contribute to the detection of various diseases through Computer-Aided Diagnosis (CAD) systems. Compared to CT scans, conducting X-rays is more accessible and cost-effective. Therefore, CXR has been chosen as the diagnostic mechanism for detecting COVID-19 and pneumonia~\cite{jacobi2020portable},\cite{cxr}.
A convolutional neural network (CNN) has been utilized for image vision tasks such as classification and recognition. In medical image processing, CNNs can take Chest X-rays (CXRs) as input, assign learnable weights and biases to different aspects of the image, and classify diseases. Generally, CNNs require a large amount of data for training and testing deep learning models~{\cite{Narin_2021}}, but medical images often need more availability. In recent studies, Transfer Learning (TL) has been used to train and test deep learning models with limited datasets~{\cite{6}}. TL involves transferring the knowledge of a model trained on millions of images to another model with fewer images, such as medical images. Therefore, the proposed model combines TL with ResNetV50~{\cite{ResNet}} and Inceptionv3~{\cite{inception}}. These two models are pre-trained on a large dataset (ImageNet), and their knowledge is transferred to the proposed model.

In deep learning-based CNNs, each layer of the model contains distinct information in terms of feature maps. The model's final (dense) layer consists of dense features used for classification and segmentation tasks. However, as we progress from the initial to the dense layer, some features may be lost. To overcome this, fusion methods~{\cite{ultra-sound}, {\cite{fang2021novel}}} have gained popularity in research, enabling the utilization of all learned information to enhance model performance. Nonetheless, fusing feature maps becomes challenging when they have different sizes in width and height. Additionally, a model-wise fusion of different architectural types presents further difficulties. In this paper, we address these issues by proposing a novel approach that fuses TL with pre-trained models ResNetV50 and Inceptionv3, leveraging their knowledge to improve diagnostic accuracy. We provide three novel contributions in this paper, by:

\begin{enumerate}
\item Introducing a novel multilayer multimodal fusion model to enhance the classification accuracy of lung diseases and empirical evaluation is conducted with potential subsidiary proposed models.
\item Developing an independent feature map transformation module (FDSFM) to address the issue of variable-sized feature maps generated across multiple layers.
\item Providing a larger dataset (Cov-Pneum) for X-ray images by processing and merging multiple publicly available datasets and hence, evaluation of the performance of multiple state-of-art research models on this dataset.
\end{enumerate}

The rest of the paper is organized as follows: Section~{\ref{2}} discusses a deep-learning approach for chest X-ray image classification. In Section~{\ref{3}}, the proposed multilayer multimodal is discussed along with the experiment details. The discussion about the dataset, preprocessing and training procedure is given in Section~{\ref{4}}. Section~{\ref{5}} considers the empirical evaluation of the model, including subsidiary proposed models and extensive experimentation. Section~{\ref{6}} discusses the quantitative and qualitative analysis of the outcome of the proposed model. Discussion on results generated by the model for the different scenarios and future directions of the proposed research is presented in Section~{\ref{7}}, and we conclude the paper in Section~{\ref{8}}.

\section{Related Work}\label{2}

Deep learning algorithms have shown a significant role in image processing and classification of lung diseases such as tuberculosis (TB), bacterial Pneumonia (BP), viral Pneumonia (VP) and COVID-19, from CXR images~\cite{{subramanian2022review},{Trans-edl}}. However, designing an effective lung disease detection model at both feature selection and classification levels is challenging. This section summarised and discussed multiple CNN architectures, parameters, performances, advantages, and limitations. 

 \textbf{Classical CNN architecture:} LeCun et al.~\cite{LeNet} introduce the LeNet model, which is one of the earliest pre-trained models of simple CNN with 5 layers. Further, Krizhevsky et al.~\cite{AlexNet} propose an AlexNet model with 8 layers of CNN architectures. However, these deep learning models suffer from extracting deep features because of the shallow network. Simonyan and Zisserman~\cite{VGG} introduce a new type of CNN architecture (known as the VGG model) which consists of a fixed size 3$\times$3 layer with stride 1 (10 layers). Further, multiple variants of VGG models, such as VGG-11, VGG-16, and VGG-19 with more layers, have been utilised with a small-size convolution filter to achieve high performance.

\textbf{Transfer learning and fine-tuning based CNN architecture:} Gour and Jain~\cite{mahesh} proposed a 30-layer CovNet30 deep learning model by stacking sub-models of the pre-trained network VGG19 using logistic regression. The model runs on the COVID-19 CXR dataset of 2764 chest X-ray images and achieves 93\% accuracy for class-3 classification. Rahimzadeh and Attar~\cite{RAHIMZADEH2020100360} developed a CNN model by concatenating the two pre-trained models (i.e., Xception and ResNet50) for 3-class classification (i.e., COVID- 19, Pneumonia and general). This network was trained and tested on 11302 chest X-ray images and achieved 91.4\% accuracy. However, these models have high false detection rates because of the loss of features when traversing from shallow to deep layers, a significant concern in such neural networks.

A deep learning-based ensemble multimodal was implemented by Deb et al.~\cite{sagar} where 4 pre-trained networks are concatenated to form the model. This model achieved an accuracy of 88.98\% for 3-class classification and an accuracy of 98.58\% for 2-class classification. In addition, the first graph-based COVID-19 classification ResGNet-C model is presented by Xiang et al.~\cite{YU2021592}, integrating graph knowledge into the classification task. They also gave a ResNet101-C model based on a pre-trained network  ResNet101 that extracts the deep features of lung diseases. The graph is constructed using the Euclidean distance from these features and generating the prediction result. A time-efficient generalized model with residual separable convolution block improves the performance of the basic MobileNet model in~\cite{tangudu2022covid}. In addition, Kaya and  Gürsoy~\cite{kaya2023mobilenet} classified CXR images into COVID-19, Normal, and Pneumonia using transfer learning on the pre-trained network MobileNetV2. They Developed an enhanced fine-tuning approach that minimized the data loss and increased the number of transfer features.

\textbf{New CNN architectures:} Khan et al.~\cite{CoroNet} proposed a CoroNet model based on Xception pre-trained model with one flattened layer, one dropout layer, and two fully connected layers. This model achieved 89.6\% accuracy for 4-class cases (COVID vs Pneumonia bacterial vs Pneumonia viral vs Normal) and 95\% accuracy for 3-class classification (COVID vs Pneumonia vs Normal). Ozturk et al.~\cite{DarkCovidNet} proposed DarkCovidNet  CNN architecture of 17 different filter-size convolutional layers. This model provides 98.08\% accuracy for binary and 87.02\% for multiple-class classification. A Covidnet model is developed by Wang et al.~\cite{Covidnet} that presents a lightweight projection-expansion-projection-extension (PEPX) design to increase prediction accuracy while reducing computation complexity. A human-machine collaborative design strategy inspires this architecture. However, the efficacy of this model cannot be generalized due to validation on the small dataset. 

Moreover, some of the models developed the model related to other lung diseases such as tuberculosis (TB), bacterial Pneumonia (BP), and viral Pneumonia (VP). In~\cite{XU202196}, Yujia et al. presented a two-stage segmented based classification model MANet to classify lung diseases into five classes (COVID-19, Normal, tuberculosis (TB), bacterial Pneumonia (BP), and viral Pneumonia (VP)). They also presented the mask attention mechanism (MA) by defining the attention maps for all features before feature extraction in CNNs via a segmentation model trained at the first stage to reduce the computational cost and distribute the attention of CNNs in the segmented lung regions.

Despite CNN, some researchers used other techniques to develop the lung disease classifier. Pathak et al.~\cite{pathak} developed a deep bidirectional extended short-term memory network with a mixture density network model for classifying COVID-19 disease. They also developed a technique to optimize the hyperparameter of the model. In addition,  Li and Xu developed a FedFocus model using a federated learning framework with a dynamic focus for COVID-19 detection on CXR images~\cite{zheng}. They set parameter aggregation weights by training loss to enhance the training efficiency. This work aimed to improve the precision and stability of the COVID-19 detection model. 

In contrast, Mohagheghi et al.~\cite{Mohagheghi} presented a three-module deep learning model for classifying COVID-19, healthy, and Pneumonia cases. CNN and content-based medical image retrieval algorithms provided the percentage of lung infection and displayed the GGO locations. Here, diagnosis of lung diseases is performed by CNN, which is the combination of VGG and hash function. If the patient is positive for COVID-19, segmentation of the abNormalities and lung lobes of the corresponding CT image are performed with the CT involvement score calculated. Similarly, Chaudhary et al.~\cite {choudhary2022deep} developed an only weight transfer method using CT images to identify COVID-19 in which they transfer the knowledge of VGG16 and ResNet34 pre-trained models. It contributed to deploying heavy ML programs in lightweight devices. In addition, Wang et al.~\cite{wang2023elucnn} developed a mobile app ELUCNN to diagnose COVID-19. They succeeded tremendously with a ten-layer CNN model with the exponential linear unit (ELU). Similarly, the MDA strategy was used in data preprocessing.

A different approach presented by Ieracitano et al.~\cite{IERACITANO2022202} based on fuzzy logic deep learning CovNNet model to classify the CXR images of patients with Covid-19 Pneumonia and with interstitial Pneumonia not related to Covid-19. CovNNet extracts the deep features and combines them with fuzzy images generated by the fuzzy edge detection algorithm. It improved the accuracy of the model up to 81\%. Zhou et al.~\cite{zhou} enlightened the issue in fine-tuning the COVID detection CNN model, in which the small dataset and the significant domain shift are the main issues. They developed a novel domain adaptation method semi-supervised open set domain adversarial network for these. It aligns a data distribution to general data space. In the presented work, results are localized by gradient-weighted class activation mapping. Dhere and Sivaswamy~\cite{Dhere} presented a multi-scale attention residual learning architecture to classify chest x-ray images into COVID-19, Pneumonia and healthy cases. With the new loss function, the Conicity loss and hierarchical method model achieved 93\%, 96.28\%, and 84.51\% accuracy for the three datasets. This work used the DenseNet model to separate pneumonia cases from normal cases and then used MARL architecture to discriminate between COVID and non-COVID Pneumonia cases.

\textbf{Fusion-based CNN architecture:} The performance of a CNN model relies on the features extracted by its layers. However, these features can be lost when transitioning from the initial to the dense layer. To address this limitation, several models have been proposed to fuse these features by combining multiple models. For instance, Muhammad and Hossain~\cite{ultra-sound} introduced a CNN model that applied layer-wise fusion for detecting COVID-19 and pneumonia. This fusion technique improved accuracy by 5.9\% compared to the model without fusion. Similarly, Ilhan et al.~\cite{ilhan2022decision} fused seven CNN architectures at the feature level and classified CXR images into COVID-19, None, and Pneumonia categories using a majority voting scheme. Waisy et al.~\cite{al2020covid} fused the features of two pre-trained models, ResNet34 and HRNet, using CXR image scores to identify COVID-19 disease. Khan et al.~\cite{KHAN2021111} presented an image scene geometry recognition model that utilized feature fusion and score-level fusion strategies to combine low-level handcrafted features with deep CNN multi-stage features, resulting in a 12.21\% improvement in accuracy. Moreover, Fang et al.~\cite{fang2021novel} proposed a multi-stage feature fusion network for COVID-19 detection by enhancing low-level feature maps. They employed ResNet-18 with a multi-stage feature enhancement module to extract and enhance these features. The development of an effective layer-wise fusion model could address the problem of feature loss and facilitate feature reusability.

Tabik et al., proposed a COVID-SDNet methodology~\cite{Tabik} that includes a fusion of segmentation, data transformation, data augmentation, and a suitable CNN model to classify the CXR images into Normal, mild, moderate, and severe COVID-19. The segmentation-based cropping method enhanced the preprocessed images by eliminating irrelevant information, improving the classification results. However, this work is limited to becoming a generalized model due to the small dataset size. Shi et al.~\cite {shi} presented an explainable attention-transfer classification model based on the knowledge distillation network structure for diagnosing COVID-19. The attention transfer direction goes from the teacher network to the student network, in which the teacher network extracts global features and generates attention maps for infection regions. The student network extracts the irregular-shaped lesion regions to learn the discriminative features. Moreover, image fusion is also performed to combine the attentive knowledge and essential information of the original input. Guest editorial contribution by Meena and Arya~{\cite{meena2023multimodal}} enlightened the various challenges and future aspects in the multimodal integration of inputs/outputs. They provided comprehensive studies of different multimodal approaches in lightweight environments, enhancing the interpretability and accuracy of classification systems. The limitation of this work is the restricted subjective dataset of humans on which the model interpretability accuracy can be checked.

\textbf{CNN architecture for other scenarios:} Another work done on COVID-19 is detecting the spread of COVID-19. In this direction, Alassafi et al.~\cite{ALASSAFI2022335} presented a model for predicting the trend of the COVID-19 outbreak. This model analyzes data from three countries Malaysia, Morocco, and Saudi Arabia. The given model uses recurrent neural network (RNN) and long short-term memory (LSTM) networks. In contrast of this, Quilodrán-Casas et al.~\cite{QUILODRANCASAS202211} introduce the SEIRS model based on LSTM and GAN with a twin digital network to predict the COVID-19 outbreak in the UK. Other than this, Wu et al. ~\cite{wu2021evolving} used Extreme Learning Machine (ELM) at the place of the classification layer in the CNN model and used a sine-cosine algorithm to tune the ELM’s parameter. In ~\cite{TAN202236}, Tan et al. introduce a multi-model model trained on CXR and CT images. The Functional AI model test CXR images and generate classification result in terms of AUC (from 0.89 to 0.93 for binary classification). 

A deeper analysis of the above state-of-the-art models shows that deep learning models suffer from high false detection rates and low accuracy. The model also suffered from overfitting due to the limited availability of labelled medical images. Therefore, facing these limitations, we introduced a significantly sized dataset (Cov-Pneum) to overcome data sparsity issues. Similarly, the proposed Multilayer multimodal fusion model enhanced the dense features by fusing feature maps of different layers sizes, reducing the false detection rate and enhancing the classification accuracy.

\section{Proposed Multilayer Multimodal Fusion Model} \label{3}

\begin{figure*}[t]
\centerline{\includegraphics[width=\linewidth,height=9cm]{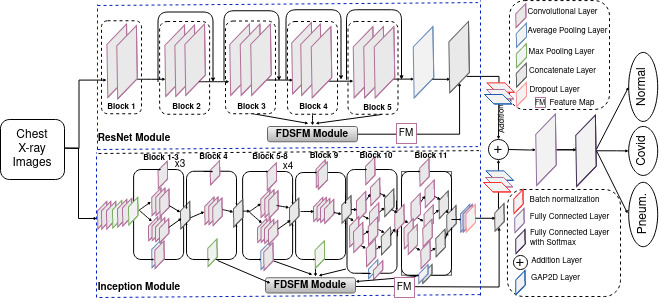}}
\caption{The architecture model of the multilayer multimodal fusion incorporates the InceptionV3 and ResNet50V2 models. In the InceptionV3 model, blocks 1 to 11 consist of layers with different-sized filters (1$\times$1, 3$\times$3, and 5$\times$5) in parallel mode. Similarly, the ResNet50V2 model includes layers from blocks 1 to 5 with different-sized filters (1$\times$1 and 3$\times$3). The FM box represents the extracted feature maps generated from different layers of their respective networks.
}
\label{fig2}
\end{figure*}

 \begin{figure}
\centerline{\includegraphics[width=\columnwidth,height=5cm]{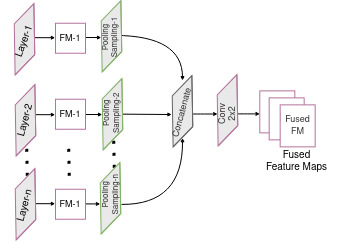}}
\caption{The layered diagram illustrates the novel module called Fusion of Different-Sized Feature Maps (FDSFM). The size transformation is achieved through pooling sampling and the Conv 1$\times$1 layer, enabling effective fusion of feature maps.}
\label{fig3}
\end{figure}

The proposed approach incorporates layer-wise fusion of multiple layers and model-wise fusion of different models. We have employed the early fusion method, which combines modalities before the classification task. Our model is built on the ResNet50V3 and InceptionV3 deep learning models, renowned for their speed and accuracy in various lung disease detection tasks \cite{fu2023pka}, \cite{cannata2022deep}, \cite{srivastava2022covixnet}. These models excel at extracting deep features through various combinations of skip and residual connections. While ResNet50V3 excels at handling deep networks and addressing the vanishing gradient problem, InceptionV3's modules facilitate multi-scale feature extraction. The synergy of these models forms a robust framework for precisely identifying intricate patterns and subtle anomalies in CXR images. Therefore, we employ ResNet50V3 and InceptionV3 in this work. Detailed explanations of our model architecture, multilayer fusion, FDSFM module, and multi-modal fusion are provided in the subsequent sections.

\subsection{Model architecture}

The proposed architecture for multilayer multimodal fusion is illustrated in Fig.~{\ref{fig2}}. Two modules, namely ResNet and Inception, represent the pre-trained ResNet50V2 and InceptionV3 models, respectively. They are utilized for feature extraction in the form of feature maps. These feature maps are then resized using the fusion of different-sized feature maps (FDSFM) module is illustrated in Fig.~{\ref{fig3}}. The FDSFM module consists of a max pooling layer for each feature map, a concatenation layer, and a 1$\times$1 convolutional (Conv) layer. This approach enhances the model's ability to learn discriminative features from CXR images.

After the multilayer fusion, a BatchNormalization layer, a 1$\times$1 Conv layer, and a Global Average Pooling 2D (GAP2D) layer are applied. These additional layers contribute to improving the generalization, convergence, and computational efficiency of the proposed model. Furthermore, multimodal fusion is performed using the addition operation to capture the distinctive characteristics of CXR images, such as ground-glass opacity and pleural effusion. To mitigate overfitting, a dropout of 30\% is employed, and the flattened output is passed through a feed-forward neural network consisting of two dense layers. The final layer of the network employs softmax activation for classification. Table~{\ref{tab:discr-table}} presents the detailed architecture of the proposed model, including the number and types of layers, as well as the output size and parameters of the model.

\begin{table}[]
\caption{The proposed Multilayer multimodal fusion model consists of layers with different types, kernel sizes, output sizes, and parameters.}
\label{tab:discr-table}
\begin{tabular}{@{}llc@{}}
\toprule
\multicolumn{1}{c}{Layer} & O/p / Kernal size & \multicolumn{1}{l}{\#Param} \\ \midrule
Pre-trained ResNet & \multicolumn{1}{c}{-} & - \\
Pre-trained Inception & \multicolumn{1}{c}{-} & - \\
ResNet\_Layer\_FM1 & (None, 28, 28, 128) & - \\
ResNet\_Layer\_FM2 & (None, 7, 7, 1024) & - \\
ResNet\_Layer\_FM3 & (None, 14, 14, 512) & - \\
ResNet\_Layer\_FM4 & (None, 28, 28, 256) & - \\
Inception\_Layer\_FM1 & (None, 5, 5, 384) & - \\
Inception\_Layer\_FM2 & (None, 5, 5, 448) & - \\
Inception\_Layer\_FM3 & (None, 5, 5, 384) & - \\
Inception\_Layer\_FM4 & (None, 5, 5, 448) & - \\
max\_pooling2d\_36 & K : 4 $\times$ 4 & 0 \\
max\_pooling2d\_39 & K : 2 $\times$ 2 & 0 \\
max\_pooling2d\_37 & K : 4 $\times$ 4 & 0 \\
concatenate\_18 & (None, 7, 7, 1920) & 0 \\
concatenate\_24 & (None, 5, 5, 1664) & 0 \\
concatenate\_19 & (None, 7, 7, 3968) & 0 \\
concatenate\_25 & (None, 5, 5, 3584) & 0 \\
batch\_norm\_112 & (None, 7, 7, 3968) & 15872 \\
batch\_norm\_115 & (None, 5, 5, 3584) & 14336 \\
conv2d\_102 & K : 2 $\times$ 2  & 8128512 \\
conv2d\_105 & K : 2 $\times$ 2 & 7342080 \\
global\_avg\_pool2d\_4 & (None, 2048) & 0 \\
global\_avg\_pool2d\_7 & (None, 2048) & 0 \\
lambda (Add\_func.) & (None, 2048) & 0 \\
dense\_2 & (None, 256) & 524544 \\
dense\_3 & (None, 3) & 771 \\ \midrule
\multicolumn{3}{l}{Total params: 61,393,699} \\
\multicolumn{3}{l}{Trainable params: 16,011,011} \\
\multicolumn{3}{l}{Non-trainable params: 45,382,688} \\ \bottomrule
\end{tabular}
\end{table}

\subsection{Multilayer fusion}
We have designed a multilayer fusion model by performing fusion on the layers of the ResNet50V2 and InceptionV3 models. The ResNet50V2 model utilizes a residual architecture and introduces two $3 \times 3$ convolutional layers with 16, 32, 64, and 128 filter depths. On the other hand, the InceptionV3 model is based on the inception architecture, where parallel layers extract deep features. Specifically, different-sized filters ($1 \times 1$, $3 \times 3$, and $5 \times 5$) are applied in parallel mode to achieve optimal performance. Initially, a Conv $1 \times 1$ layer is used before the Conv $3 \times 3$ and Conv $5 \times 5$ layers to reduce the number of parameters. The outputs of these filters are concatenated and passed through a pooling layer with a size of $3 \times 3$. To fuse dense features and prevent feature loss, we consider the Conv layer from the middle module of the models.

The size of a feature map is determined by its length, width, and channel, and only feature maps with the same dimensions can be fused. To minimize the transformation overhead and information loss, we have considered a selection layer that can choose feature maps of the same length and width generated from a specific layer.However, feature maps of the same size cannot be obtained from layers. Therefore, Section~\ref{fdsfm} proposes a novel fusion model that appropriately resizes the feature maps. Subsequently, the resulting feature maps from each model are fused at the concatenation layer. At the intermediate layer, we combine these feature maps using Equation (\ref{eq1}).

 \begin{equation}
 \label{eq1}
     X'_{Concat} = x'_{1} \cup  x'_{2} \cup  x'_{3} \cup \ldots \cup   x'{_i} \cup \ldots \cup  x'{_n}
 \end{equation}
 Where $x'_{n}$ is the features maps of $n^{th}$ layer and $X'_{Concat} $ is the outcome of the fused features map.

\subsection{Fusion of different sized feature maps}\label{fdsfm}

The fusion of multiple layers generated feature maps is still challenging due to their variable sizes. To encounter this issue, we designed a novel fusion of different-sized feature maps (FDSFM) method as illustrated in Fig. {\ref{fig3}}. It first reduces the size of the feature map to the optimal size where all feature maps can be fused efficiently. We applied MaxPooling2D layer of filter size \textit{f} and strides of \textit{s} to downsample the input. We used Eq. (\ref{eq3}) to reduce the features map of $X^{(W \mathrm{x} L \mathrm{x} C)}$ into the appropriate size. Where \textit{L}, \textit{W} and \textit{C} represent length, width and number of channels, respectively. Afterwards, for effective feature mapping, we applied pooling sampling Conv ($1 \times 1$) for channel size adjustment.

 \begin{equation}
 \label{eq3}
X^{(W' \times L' \times C')} = X^{\Big(\big((W-f+1)/s\big) \: \times \:\big((L-f+1)/s\big) \:\times \:C\Big)}
 \end{equation}

\subsection{Multimodal fusion}
 Similar multilayer fusion model, we designed a multimodal fusion model by performing the fusion between ResNet50V2 and InceptionV3 models at the model level. From the multilayer fusion model, we received two different feature maps at the concatenate layer of each model. To remove the feature map size restrictions, each feature map passes through the Global Average Pooling 2D (GAP2D) layer. We designed and performed model-wise fusion by using Eq.(~\ref{eq2}).
 \begin{equation}
 \label{eq2}
     X'_{Add} = X'_{Concat1} \odot X'_{Concat2} 
 \end{equation}
 
Where $X'_{Concat1}$ and $X'_{Concat2} $ are the outcome of layer-wise feature maps of the individual model. We performed element-wise addition $(\odot)$ on these features maps to generate final fused features maps. We noticed that addition fusion performed better as compared to concatenate fusion. 

When the model was developed, we trained this proposed model on a given dataset of N$_{s}$ training samples, where  N$_{s}$=(X,$\mathfrak{L}$).We denote input CXR images by X = $\{ x_1, x_2, x_3, \ldots, x_n\}$ and their corresponding true labels by  $\mathfrak{L}$ =$\{ t_1, t_2, t_3, \ldots, t_n\}$. We denote the total number of classes that are available any time in the model by $t_i$, therefore $t_i \in \{0,1,2\}$ where 0, 1, and 2 represent COVID-19, Pneumonia, and Normal class labels, respectively. The total number of layers in model given as L=$\{ l_1, l_2, l_3, \ldots,  l_n\}$ and  corresponding features maps are represented as $X'=$ \{$x_1', x_2', x_3', \ldots, x_i'\ldots, x_n'\}$,  where each features map is given by $x_i'\in \Re^{W,L,C}$   with W (width), L (length), and C (number of channels).
 
The working functionality of the proposed model can be shown by the function $f^w : X \rightarrow \mathfrak{L}$ that is parameterized by weight w. We performed an intermediate process before generating of predicted class label ($\mathfrak{L}$). The predicted class score is calculated by $f (X_{i}; w)$ and the outcome of this is a real number from $-\infty$ to $\infty)$. Now to restrict this output score,  we convert it into prediction probability and applied the logistic function softmax as given in Eq.~(\ref{eq5}).
 \begin{equation}
 \label{eq5}
 p\left(l_{i} \mid\left(f\left(X^{j}; w\right)\right)\right)=\frac{e^{f_{i}\left(X^{j}; w\right)}}{\sum_{i=1}^{n} e^{f_{i}\left(X^{j} ; w\right)}}
 \end{equation}
 
Further, we trained the proposed network model to minimize the loss and the desired output was  generated by assigning the highest probability to the correct output labels as: 
\begin{equation}
 \label{eq6}
w^{*}=\arg \min _{w} \sum_{j=1}^{n} \operatorname{loss}\left(t_{i} , p\left(l_{i} \mid\left(f_{i}\left(X^{j}, w\right)\right)\right)\right)
\end{equation}
Where loss is considered as categorical-cross-entropy by Eq.~(\ref{eq7}).
\begin{equation}
 \label{eq7}
\begin{split}
\operatorname{loss}\left(t_{i}, p\left(l_{i}\mid\left(f_{i}\left(X^{j}; w\right)\right)\right)\right) = \\ \sum_{i=1}^{n}-\log p  \left(l_{i} \mid\left(f_{i}\left(X^{j}; w\right)\right)\right) \cdot t_{i}
\end{split}
\end{equation}

\begin{table*}[]
\caption{Size of the dataset for all classes (COVID-19 (C), Pneumonia (P), Normal (N)) that are used to validate the state-of-art CNN models and their dataset comparison with our proposed Cov-Pneum dataset.}
\label{tab:my-result}
\centering
\begin{tabular}{@{}lllll@{}}
\toprule
\multirow{2}{*}{Dataset} & \multicolumn{4}{l}{Size of CXR images} \\
 & C & P & N & Sum \\ \midrule
DarkCovid-Net~\cite{DarkCovidNet} & 125 & 500 & 500 & 1125 \\
CoroNet~\cite{CoroNet} & 3180 & - & 10120 & 13800 \\
CovidNet~\cite{Covidnet} & - & - & - & 13975 \\
CovidGAN~\cite{Covidgan} & 403 & - & 721 & 1124 \\
Rahimzadeh et al.~\cite{RAHIMZADEH2020100360} & 180 & 6054 & 8851 & 15085 \\
Shanjiang et al.~\cite{Trans-edl} & 473 & 5459 & 7966 & 13898 \\
\textbf{Cov-Pneum dataset} & \textbf{4296} & \textbf{5824} & \textbf{11152} & \textbf{21272} \\ \bottomrule
\end{tabular}
\end{table*}

The proposed model is then trained on the Cov-Pneum CXR dataset for training, validation and testing.

\section{Dataset, Preprocessing and Training Procedure}\label{4}
In this section, we present a step-by-step description of the data preparation, preprocessing, and training and optimization processes.

\subsection{Dataset description}

We prepared a new dataset called Cov-Pneum by processing and merging three well-known publicly available datasets from Kaggle~\cite{24,25,26,27}. The Cov-Pneum dataset consists of a total of 21,272 CXR images categorized into COVID-19 (lung infected with COVID-19 virus), Pneumonia (viral Pneumonia, non-COVID-19 Pneumonia, and COVID-19 Pneumonia infected lung), and Normal (clear lung) classes. The dataset contains 4,296 COVID-19 images, 5,824 Pneumonia images, and 11,152 Normal images. To improve the quality of the CXR images, we applied image scaling and pre-processing operations.

Table~\ref{tab:my-result} presents a comparison of the proposed Cov-Pneum dataset with other state-of-the-art datasets used in multiple studies. We will publish the Cov-Pneum dataset online and provide a link upon final approval. This will enable fellow researchers to validate their models on a significant-sized dataset and address data sparsity issues.

\subsection{Data preprocessing}
In the preprocessing stage, the CXR images were resized using bilinear interpolation to ensure a uniform size and standardized dimensions. The pixel intensities of the images were normalized within the range of (-1, 1). This ensures that the input data is appropriately scaled, leading to more effective and stable training.

Furthermore, in the preprocessing pipeline, we applied geometric transformations using shear range and zoom parameters. A shear range of 0.1 allowed us to perform affine transformations on the images, horizontally skewing them by a maximum of 10\% in either direction. Similarly, a zoom factor of 0.1 enabled us to randomly zoom the images in or out, with a maximum variation of 10\%. These transformations add variability to the dataset, enhancing the model's ability to learn and generalize from different orientations, scales, and geometries present in the CXR images.

\subsection{Training and optimization}

All experiments were conducted using pre-trained models on the ImageNet dataset~{\cite{deng2009imagenet}}. Subsequently, these models were trained on the Cov-Pneum dataset, which consists of CXR images with a size of (224, 224, 3). The supervised learning approach was employed to train the proposed model and sub-models on a total of 12,157 labeled CXR images categorized as COVID-19, Pneumonia, and Normal. Similarly, 3,219 CXR images were used for validation, and the remaining 5,896 CXR images were allocated for testing.

The proposed network performed a series of convolutional, pooling, and activation operations on the data to extract meaningful features. The output of the network was then compared to the true labels using a categorical cross-entropy loss function given in Equation ({\ref{eq7}}). The primary objective of using this loss function was to minimize the error. To optimize the learning weights of the model, the Adam optimizer~{\cite{kingma2014adam}} was employed due to its effectiveness in other classification models~{\cite{bera2020analysis}}.

The performance of the CNN model depends on several hyperparameters, including the batch size, number of epochs, and learning rate. A range of values was explored for each hyperparameter to evaluate their impact on the model's accuracy. For the learning rate, values ranging from 0.0001 to 0.1 were tested with increments of 0.01. The batch size was varied from 16 to 128, increasing by a factor of 8. Additionally, epoch values ranging from 10 to 100, with increments of 10, were experimented with. Through this systematic exploration, it was observed that a learning rate of 0.001, a batch size of 32, and 100 epochs yielded the highest accuracy for the proposed model. Various evaluation metrics, such as confusion matrix, accuracy, precision, F-1 score, recall, and ROC curve, were employed to assess and compare the performance of the model.

\section{Empirical Evaluation}\label{5}

We propose and implement additional potential models for in-depth validation and to find the efficacy of the proposed multilayer multimodal fusion model at multiple levels. Likewise, these subsidiary models also address the loss of layer features when features move from shallow layers to deeper layers. 

\subsection{Subsidiary proposed models}
In addition to the multilayer multimodal fusion model, this work proposes the three subsidiary models for in-depth comparison and evaluation.

\textbf{1. Multilayer fusion of ResNet50V2 model:} This model is based on the ResNet50V2 architecture (Fig. 1), where we employ layer-wise fusion. We extract deep features from the Block 3, Block 4, and Block 5 layers of the ResNet50V2 model. These extracted feature maps may have different sizes (length and width), which poses a challenge for fusion. To address this, we utilize the FDSFM module to transform and fuse these feature maps using Eq. (1). Furthermore, the extracted features and the output of the ResNet50V2 model (before the dense layer) are fused at the concatenate layer. A dropout layer with a rejection rate of 0.3, followed by a dense layer and a fully connected layer with softmax activation, is applied at the top of the model.

\textbf{2. Multilayer fusion of InceptionV3 model:} The same experiment is also conducted with the InceptionV3 model, employing layer-wise fusion. Deep feature maps are generated by extracting features from block 4 to block 11 layers. Fusion and size transformation of the feature maps are performed in the FDSFM module. All generated feature maps are concatenated with the last layer (before the dense layer) of the InceptionV3 model. A dropout layer with a rejection rate of 0.3, followed by a dense layer and a fully connected layer with softmax activation, is applied at the top of the model. The concatenation of deep features significantly improves the performance of the proposed multilayer fusion of the InceptionV3 model.

\textbf{3. Singlelayer multimodal fusion model:} We implemented this model using singlelayer multimodal fusion to demonstrate the difference between multilayer fusion and singlelayer fusion, as well as multimodal fusion and singlemodal fusion. Model-wise fusion was performed between the outputs of the last layers of the ResNet50V2 and InceptionV3 models. Fusion was accomplished using a point-wise addition operation, which effectively reduced the number of parameters. This operation is represented by Eq. 3. The concatenated features obtained from the fusion were then passed through a dropout layer with a rejection rate of 0.3, followed by a dense layer and a fully connected layer with softmax activation at the top of the model. As a result, the proposed model achieved remarkable success in accurately classifying the input CXRs based on the generated deep features.

\begin{figure*}
    \centering
    \subfigure[]{\includegraphics[width=0.48\textwidth]{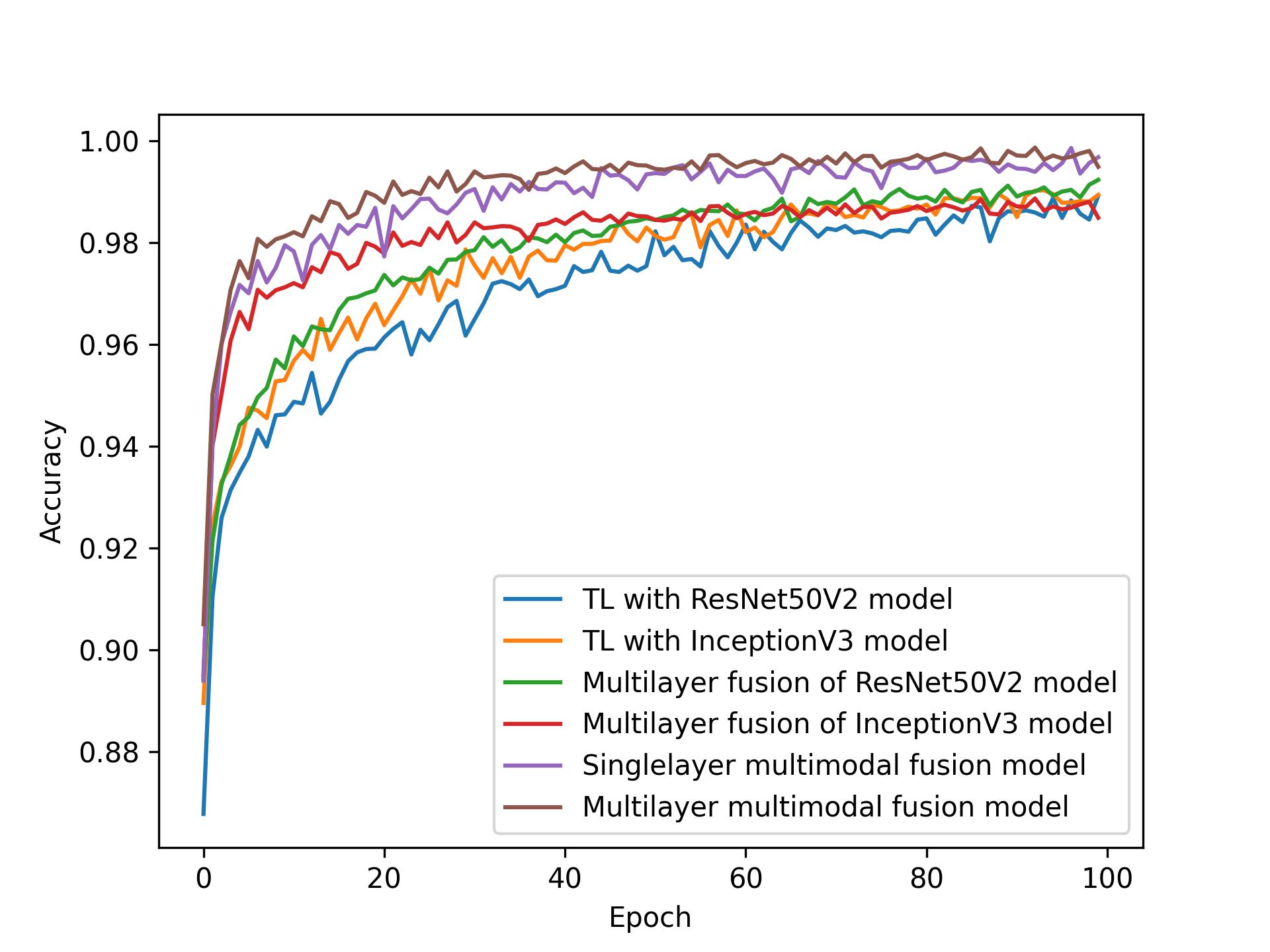}} 
    \subfigure[]{\includegraphics[width=0.48\textwidth]{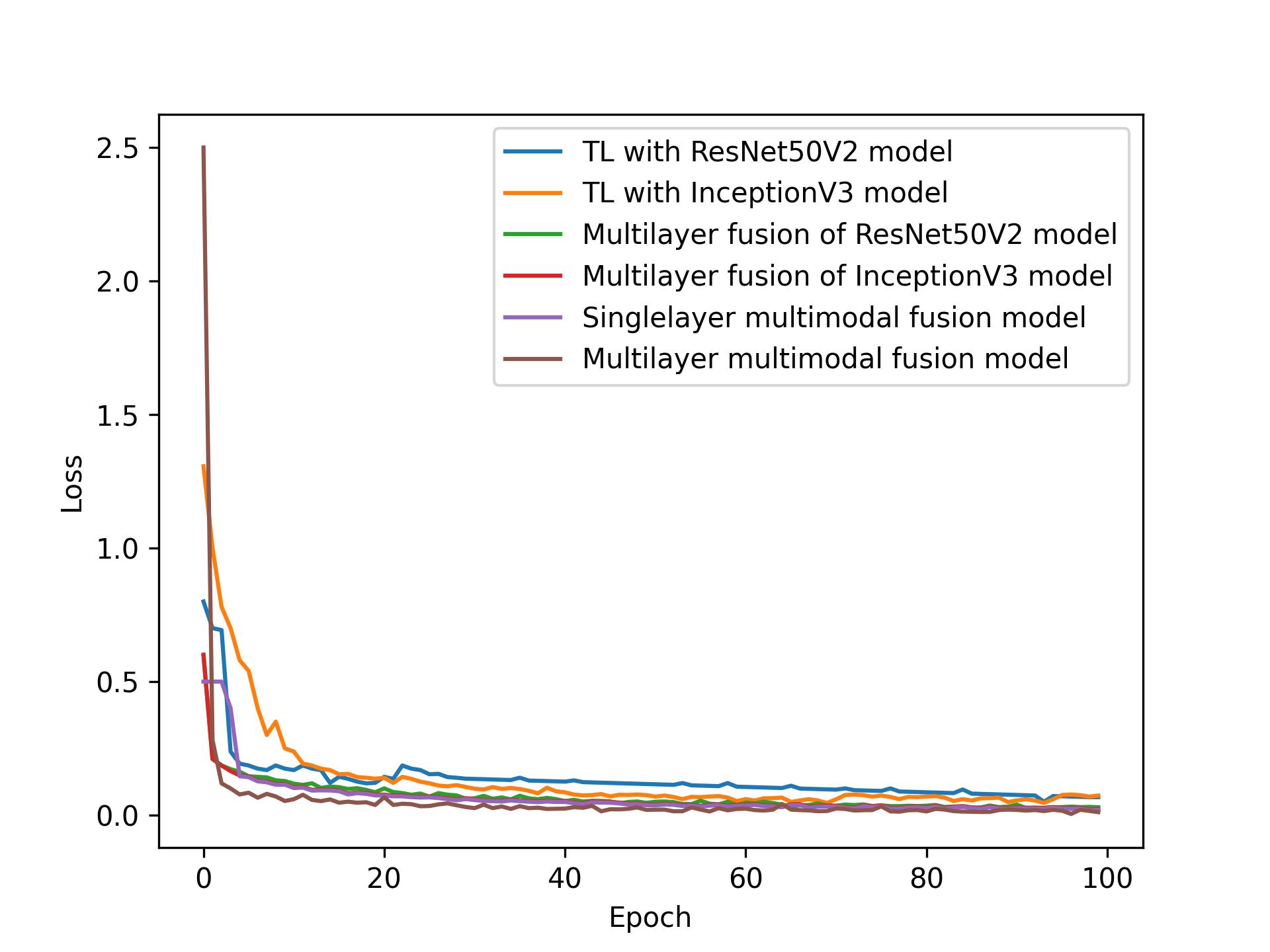}} 
    \caption{Comparison of the learning behavior of baseline models (TL with ResNet50V2~\cite{ResNet}, TL with InceptionV3~\cite{inception}) and proposed models in terms of training accuracy (a) and training loss curves at per epoch time.}
    \label{learning-behavior}
\end{figure*}

\subsection{Experiments}

Extensive experimentation has been conducted to evaluate the performance of the proposed model, which involves fusing features from different layers of the models. We implemented all four proposed models along with other state-of-the-art models on the Cov-Pneum dataset to assess the effectiveness of the proposed approach.

\textbf{Experiment 1: Singlemodal fusion v/s multimodal fusion}\\
In this experiment, we evaluate and compare the performance of singlemodal fusion (i.e., multilayer fusion of ResNet50V2, multilayer fusion of InceptionV3 model) and multimodal fusion (i.e., singlelayer multimodal fusion model, multilayer multimodal fusion model). We implemented the models discussed in Section III to demonstrate these evaluations. The experiment with layer-wise fusion is conducted on both the ResNet50V2 model and the InceptionV3 model to implement the singlemodal fusion. Similarly, the same experiment is performed on the results of these separated models to build the multimodal fusion. Furthermore, the multimodal fusion approach is applied to the output of the last layers of the ResNet50V2 and InceptionV3 models to create additional cases of multimodal fusion (singlelayer multimodal fusion model).

\textbf{Experiment 2: Singlelayer fusion v/s multilayer fusion}\\
This experiment is conducted to evaluate the performance of the singlelayer fusion model (i.e., singlelayer multimodal fusion model) with the multilayer fusion model (i.e., multilayer fusion of ResNet50V2, multilayer fusion of InceptionV3 model, and multilayer multimodal fusion model). In the singlelayer fusion approach, the output layer of each respective model is fused, while in the multilayer fusion, multiple layers of each model are combined. Consequently, we performed a substantial number of experiments on different models, ultimately achieving success by fusing the output layer of the ResNet50V2 and InceptionV3 models. Firstly, layer-wise fusion is implemented on the ResNet50V2 model to perform the multilayer fusion. Subsequently, the same experiment is also conducted on the InceptionV3 model. Furthermore, model-wise fusion is implemented on the output of these subsequent results.

\textbf{Experiment 3: State-of-art models evaluation on Cov-Pneum data-set}\\
The performance of the newly implemented model can be evaluated by comparing it with the other state-of-art methods. The system environment and dataset are the key parameters in validating the results. Therefore, we implemented the state-of-art COVID-19 and Pneumonia detection methods such as  CoroNet~\cite{CoroNet}, DarkCovid-Net~\cite{DarkCovidNet}, CovidNET~\cite{Covidnet}, Fangg et al.~\cite{fang2021novel}, Shanjiang et al.~\cite{Trans-edl}, CovidGAN~\cite{Covidgan}, TL with VGG19~\cite{TL-with-VGG19}, TL with ResNet50V2~\cite{ResNet}, TL with Xception~\cite{xception}, and TL with InceptionV3~\cite{inception} on  proposed dataset Cov-Pneum. This experiment was performed in the Jupyter Notebook environment using the Tensorflow Keras library on an Intel Xeon Z440 with 8 GB RAM and a processing speed of 3.5 GHz.

\textbf{Experiment 4: Proposed model evaluation on state-of-art datasets}\\
We conducted this experiment to evaluate the proposed model performance on state-of-art datasets. Here we considered the dataset Dark-CovidNet~\cite{DarkCovidNet} to have 1125 CXR images (COVID-19: 125, Pneumonia: 500, and Normal: 500)   and   Rahimzadeh et al.~\cite{RAHIMZADEH2020100360} dataset has 15085 CXR images (COVID-19: 180, Pneumonia: 6054 and Normal: 8851). Our proposed Multilayer multimodal fusion model is implemented on this dataset for further evaluation and validation.

\section{Results}\label{6}

We present a comprehensive analysis of the complete experiments to demonstrate the performance of our proposed work. To validate our results, we compare the performance of the proposed model with several sub-models and state-of-the-art COVID-19 and Pneumonia detection methods. It includes the proposed sub-models, such as the ResNet50V2 model's multilayer fusion of the InceptionV3 model, singlelayer multimodal fusion model, and multilayer multimodal fusion model. Whereas it includes the various state-of-the-art methods, such as CoroNet~{\cite{CoroNet}}, DarkCovid-Net~{\cite{DarkCovidNet}}, CovidNET~{\cite{Covidnet}}, Fangg et al.~{\cite{fang2021novel}}, Shanjiang et al.~{\cite{Trans-edl}}, CovidGAN~{\cite{Covidgan}}, TL with VGG19~{\cite{TL-with-VGG19}}, TL with ResNet50V2~{\cite{ResNet}}, TL with Xception~{\cite{xception}}, and TL with InceptionV3~{\cite{inception}}.

\subsection{Training performance}

The learning behavior of the models in terms of accuracy and loss curves per epoch is demonstrated in Fig.~{\ref{learning-behavior}}. It also compares the learning performance of the proposed model with different sub-models and various baseline models. As the number of epochs increases, the model shows a notable trend of achieving the highest accuracy while substantially reducing the loss. This indicates the good learning behavior of the proposed model during the training and validation phases. Interestingly, the proposed model exhibits an accuracy curve approaching one and a loss curve approaching zero, in contrast to other state-of-the-art models. This exceptional performance serves as evidence of the model's excellent training behavior.

\begin{sidewaystable}[]
\caption{Performance evaluation of state-of-art and our models on proposed Conv-Pneum dataset. Computed classification accuracy (\%) for both 3-class (3-C) and binary (2-C) class models. Evaluate individual class performance  P: Precision, R: Recall, and F-1 : F-1 Score. Here, M1 is multilayer fusion of ResNet50V2 model, M2 is multilayer fusion of InceptionV3 model, M3 is singlelayer multimodal fusion model, and M4 is  multilayer multimodal fusion model.}
\label{table:result}
\begin{tabular}{llllllllllll}
\hline
\multirow{2}{*}{\begin{tabular}[c]{@{}l@{}}  Existing vs proposed \\ models \end{tabular}} & \multicolumn{3}{c}{Covid-19} & \multicolumn{3}{c}{Pneumonia} & \multicolumn{3}{c}{Normal} & \multirow{2}{*}{\begin{tabular}[c]{@{}l@{}}  Acc\\ (3-C)\end{tabular}} & \multirow{2}{*}{\begin{tabular}[c]{@{}l@{}}  Acc\\ (2-C)\end{tabular}} \\ \cline{2-10}
\multicolumn{1}{c}{} & \multicolumn{1}{c}{P} & \multicolumn{1}{c}{R} & \multicolumn{1}{c}{F-1} & \multicolumn{1}{c}{P} & \multicolumn{1}{c}{R} & \multicolumn{1}{c}{F-1} & \multicolumn{1}{c}{P} & \multicolumn{1}{c}{R} & \multicolumn{1}{c}{F-1} &  \\ \hline
DarkCovid-Net~\cite{DarkCovidNet} & 0.97 & 0.90 & 0.94 & - & - & - & 0.98 & 0.99 & 0.98 & 87.02 & 98.08 \\
CoroNet ~\cite{CoroNet} & 0.93 & 0.98 & 0.95 & 0.84 & 0.82 & 0.83 & 0.95 & 0.93 & 0.94 & 92.45 & -\\
CovidNET~\cite{Covidnet} & 0.98 & 0.91 & 0.94 & 0.91 & 0.94 & 0.92 & 0.90 & 0.95 & 0.92 & 92.60 & 97.54 \\
CovidGAN~\cite{Covidgan} & 0.96 & 0.90 & 0.93 & - & - & - & 0.94 & 0.97 & 0.95 &  - & 94.78 \\
TL with VGG19~\cite{TL-with-VGG19} & 0.86 & 0.99 & 0.92 & 0.94 & 0.98 & 0.96 & 0.99 & 0.92 & 0.96 & 90.25 & 94.62 \\
TL with ResNet50V2~\cite{ResNet} & 0.93 & 0.91 & 0.92 & 0.95 & 0.92 & 0.93 & 0.91 & 0.94 & 0.92 & 91.74 & 94.78\\
TL with Xception~\cite{xception} & 0.95 & 0.90 & 0.92 & 0.89 & 0.91 & 0.90 & 0.87 & 0.92 & 0.89 & 92.24 & 95.23  \\
TL with InceptionV3~\cite{inception} & 0.92 & 0.94 & 0.93 & 0.98 & 0.93 & 0.96 & 0.91 & 0.98 & 0.95 & 93.87 & 95.74\\
Fangg et al.~\cite{fang2021novel} & 0.99 & 0.94 & 0.96 & 0.96 & 0.94 & 0.95 & 0.90 & 0.98 & 0.94 & 95.29 & 97.47\\
Shanjiang et al.~\cite{Trans-edl} & 0.94 & 0.96 & 0.95 & 0.93 & 0.94 & 0.93 & 0.96 & 0.95 & 0.95 & 95.69 & 98.09\\
\textbf{M1 }& \multicolumn{1}{c}{\textbf{0.90}} & \multicolumn{1}{c}{\textbf{0.92}} & \multicolumn{1}{c}{\textbf{0.91}} & \multicolumn{1}{c}{\textbf{0.99}} & \multicolumn{1}{c}{\textbf{0.93}} & \multicolumn{1}{c}{\textbf{0.96}} & \multicolumn{1}{c}{\textbf{0.84}} & \multicolumn{1}{c}{\textbf{0.97}} & \multicolumn{1}{c}{\textbf{0.90}} & \textbf{94.01} & \textbf{97.82} \\
\textbf{M2 }& \multicolumn{1}{c}{\textbf{0.92}} & \multicolumn{1}{c}{\textbf{0.91}} & \multicolumn{1}{c}{\textbf{0.92}} & \multicolumn{1}{c}{\textbf{0.99}} & \multicolumn{1}{c}{\textbf{0.94}} & \multicolumn{1}{c}{\textbf{0.96}} & \multicolumn{1}{c}{\textbf{0.83}} & \multicolumn{1}{c}{\textbf{0.97}} & \multicolumn{1}{c}{\textbf{0.90}} & \textbf{94.18} & \textbf{97.47} \\
\textbf{M3 }& \multicolumn{1}{c}{\textbf{0.97}} & \multicolumn{1}{c}{\textbf{0.93}} & \multicolumn{1}{c}{\textbf{0.95}} & \multicolumn{1}{c}{\textbf{0.99}} & \multicolumn{1}{c}{\textbf{0.98}} & \multicolumn{1}{c}{\textbf{0.99}} & \multicolumn{1}{c}{\textbf{0.95}} & \multicolumn{1}{c}{\textbf{0.93}} & \multicolumn{1}{c}{\textbf{0.94}} & \textbf{95.95} & \textbf{98.05} \\
\textbf{M4 }& \multicolumn{1}{c}{\textbf{0.96}} & \multicolumn{1}{c}{\textbf{0.98}} & \multicolumn{1}{c}{\textbf{0.97}} & \multicolumn{1}{c}{\textbf{0.99}} & \multicolumn{1}{c}{\textbf{0.97}} & \multicolumn{1}{c}{\textbf{0.98}} & \multicolumn{1}{c}{\textbf{0.94}} & \multicolumn{1}{c}{\textbf{0.96}} & \multicolumn{1}{c}{\textbf{0.95}} & \textbf{97.21} & \textbf{99.60} \\ \hline
\end{tabular}
\end{sidewaystable}

\begin{figure*}
    \centering
    \subfigure[]{\includegraphics[width=0.24\textwidth]{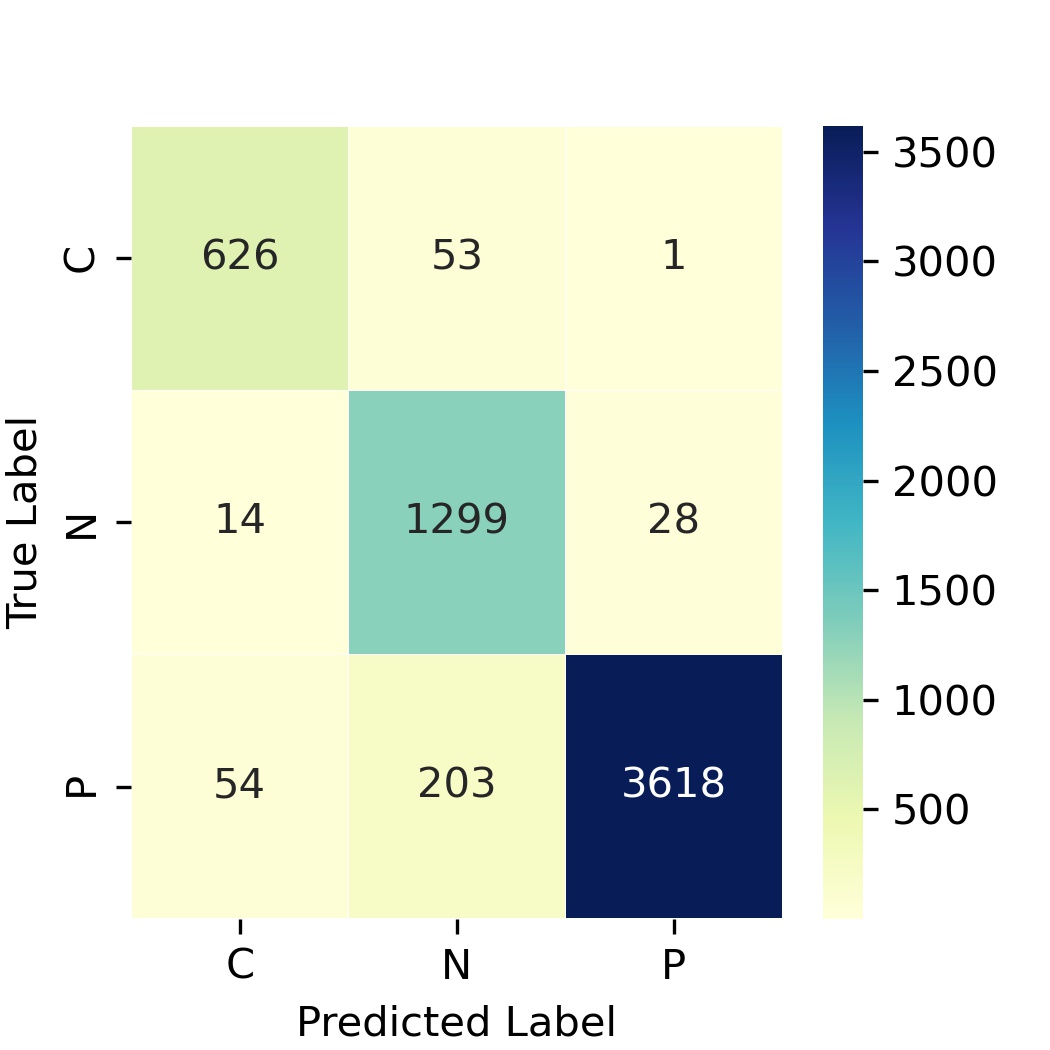}} 
    \subfigure[]{\includegraphics[width=0.24\textwidth]{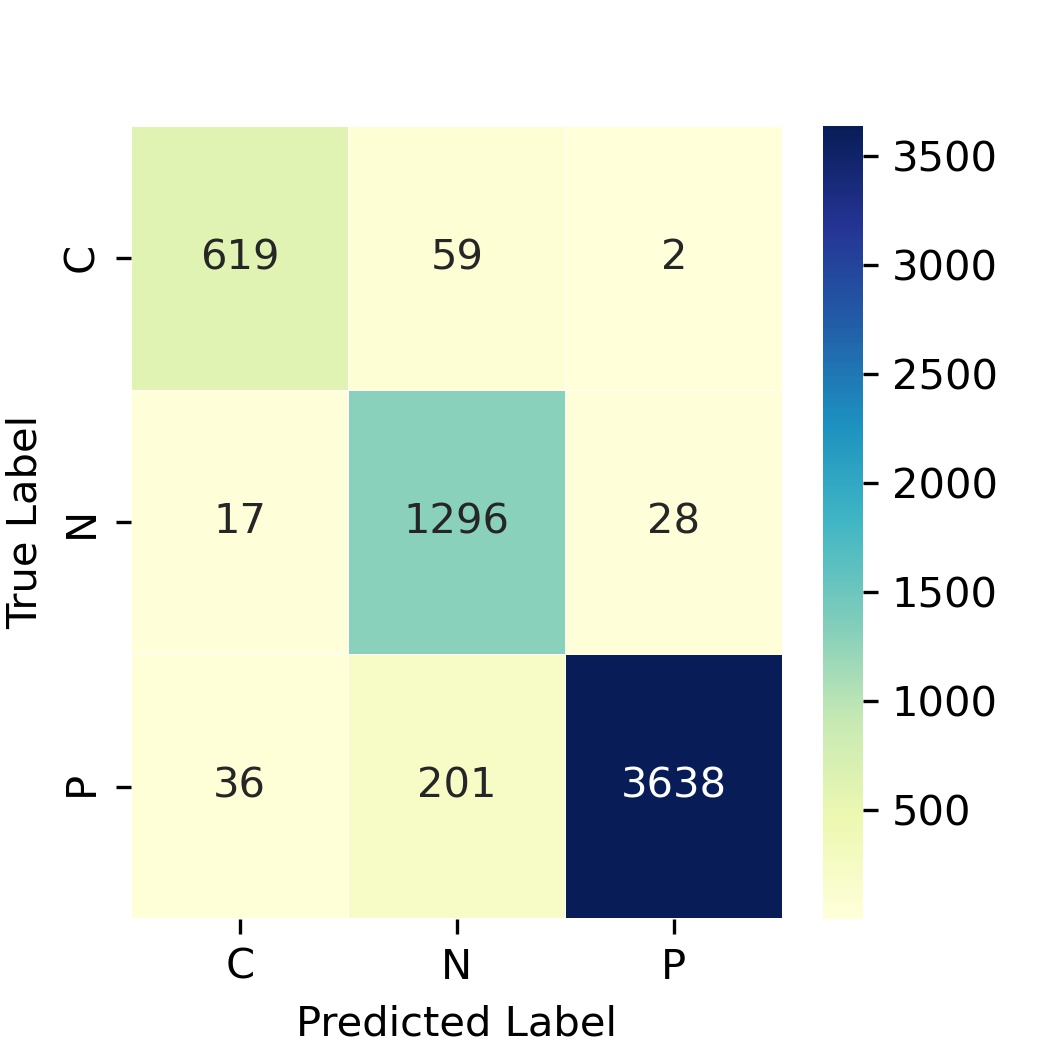}} 
    \subfigure[]{\includegraphics[width=0.24\textwidth]{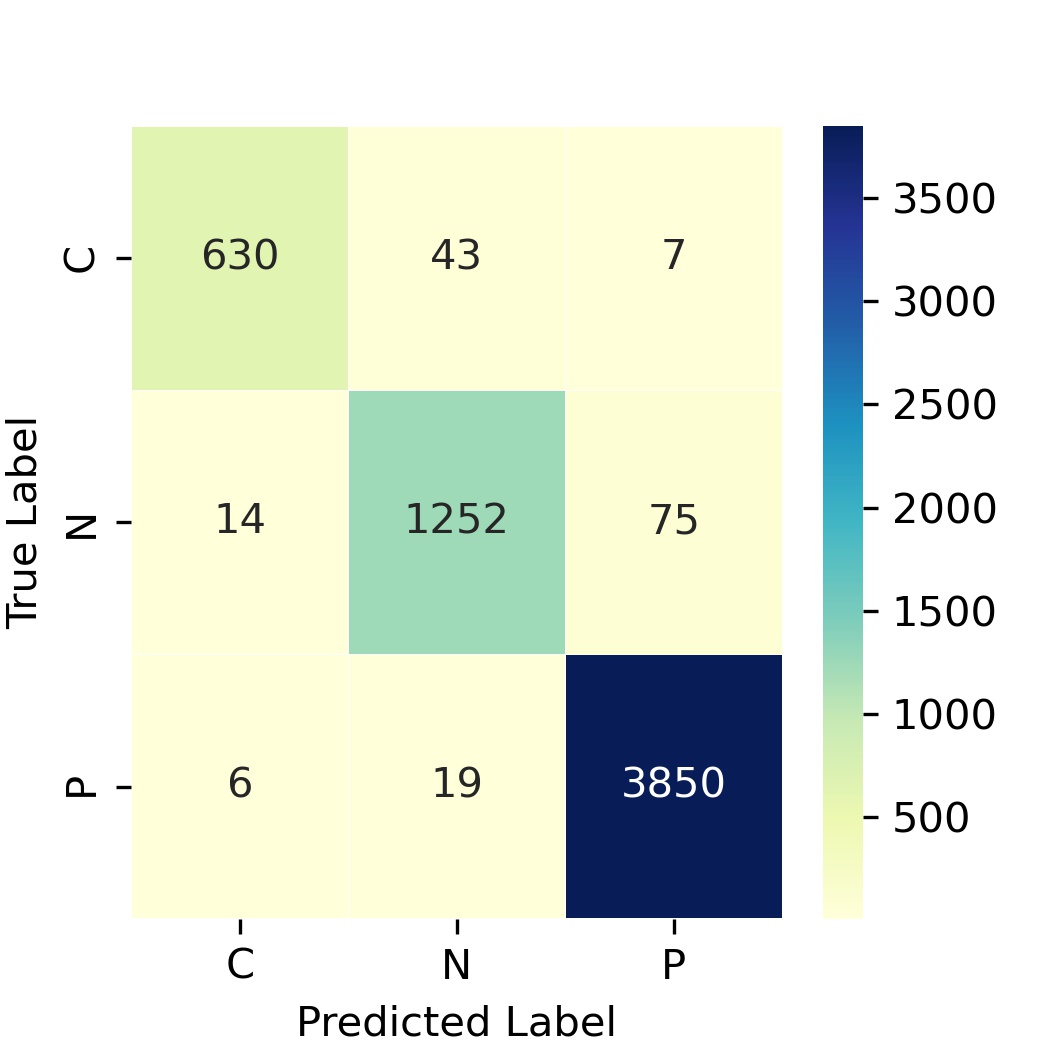}}
    \subfigure[]{\includegraphics[width=0.24\textwidth]{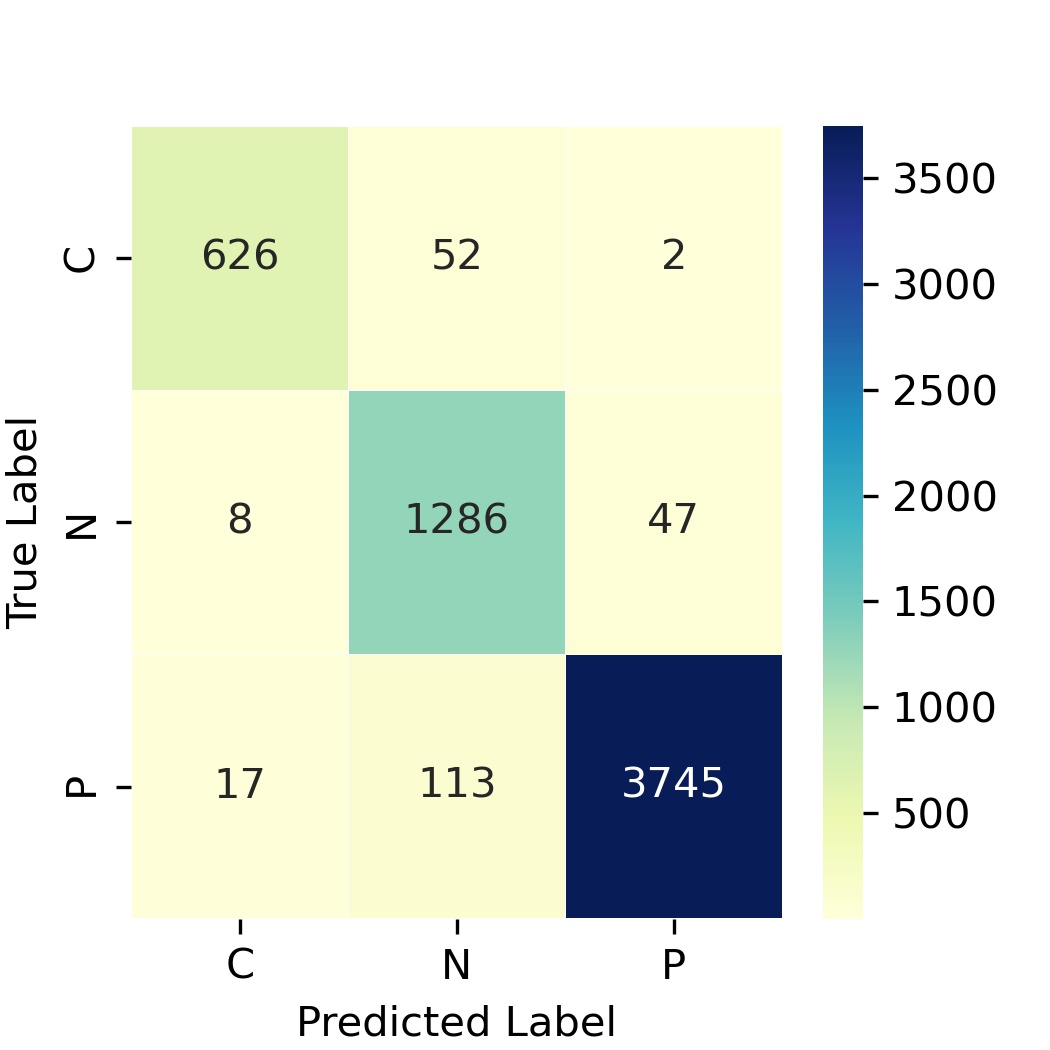}}\\
    \subfigure[]{\includegraphics[width=0.24\textwidth]{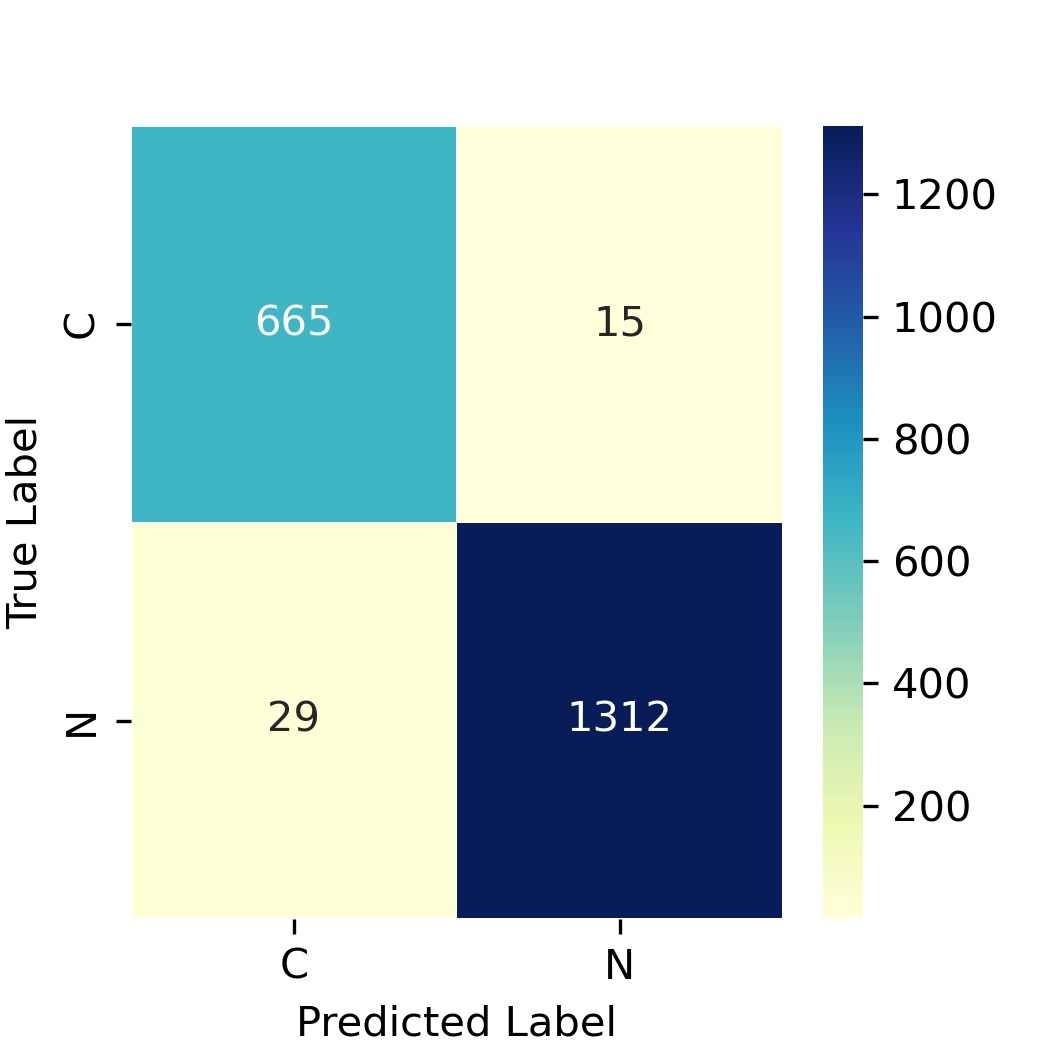}} 
    \subfigure[]{\includegraphics[width=0.24\textwidth]{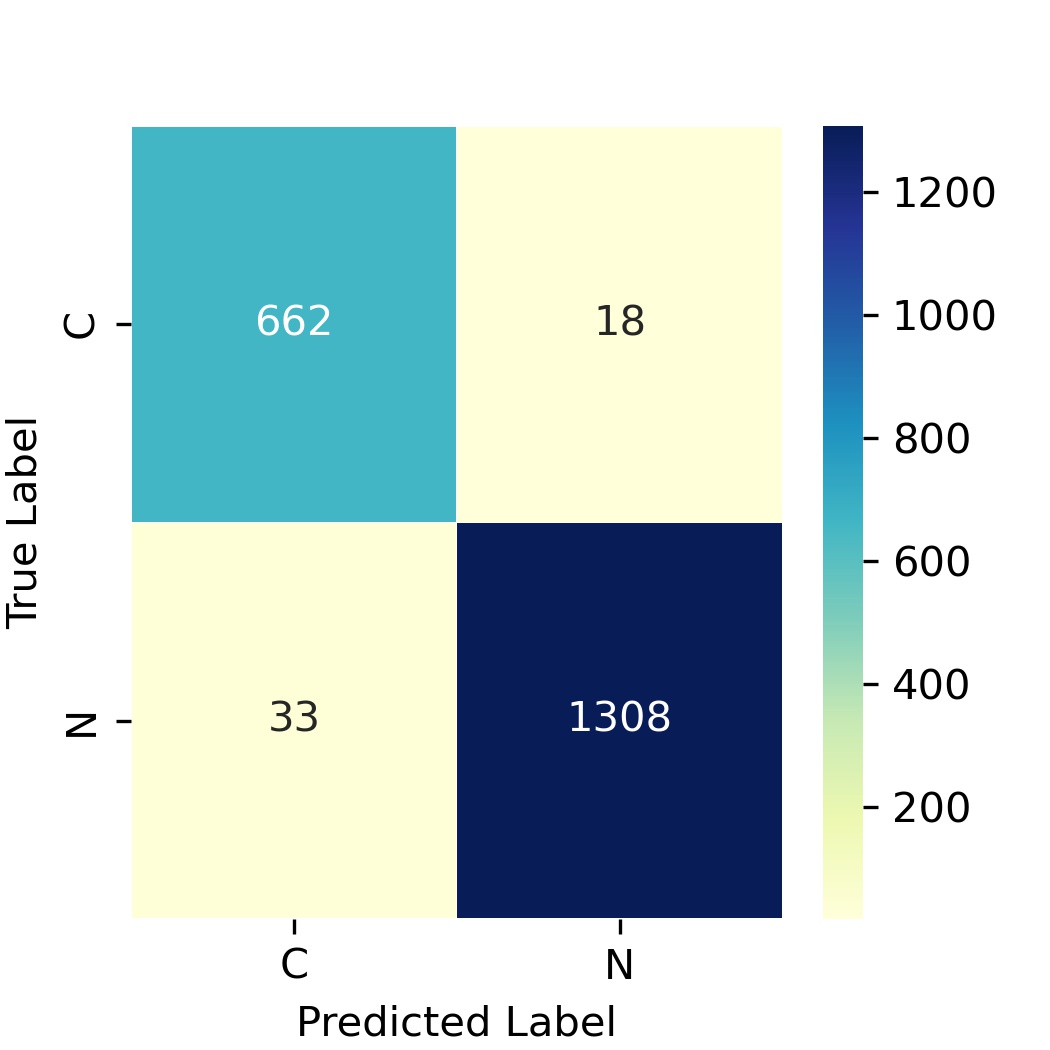}} 
    \subfigure[]{\includegraphics[width=0.24\textwidth]{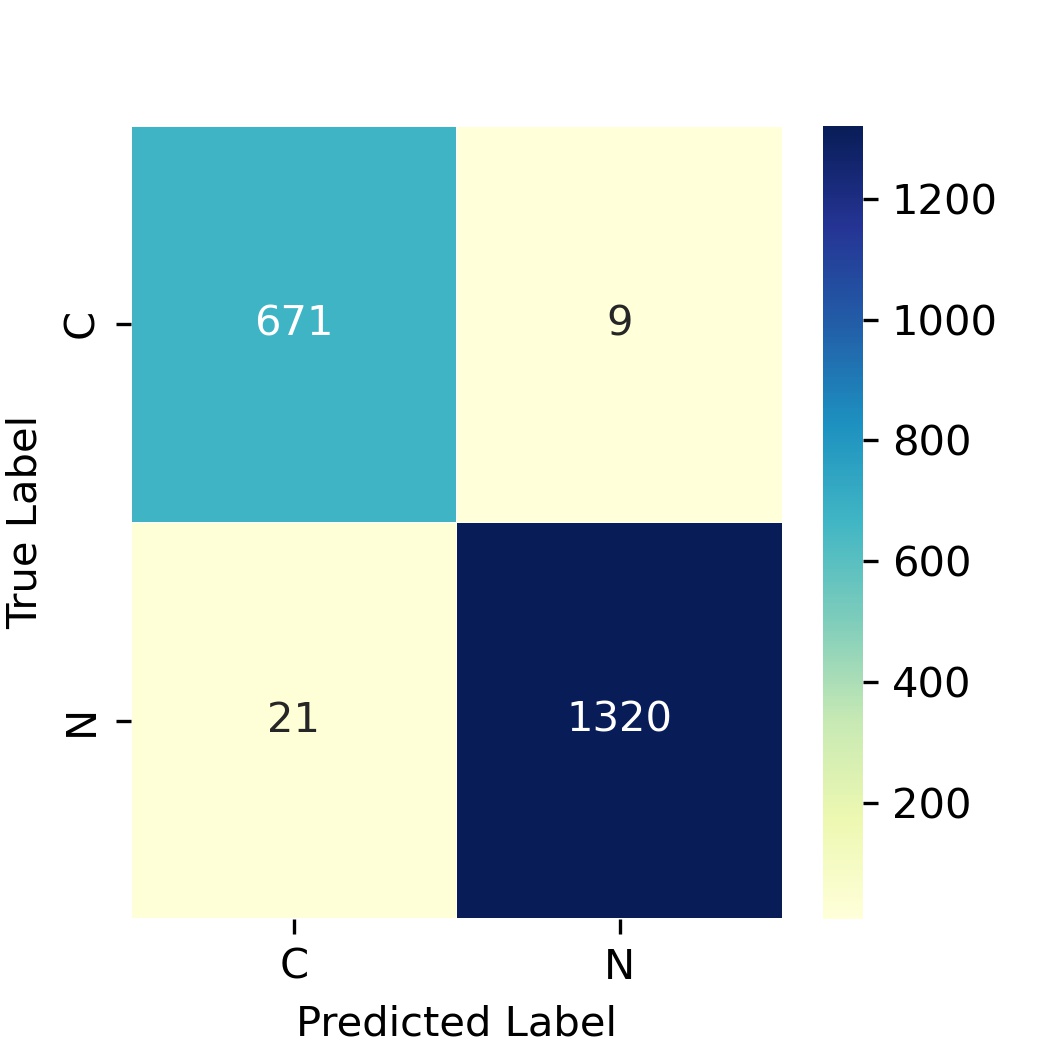}}
    \subfigure[]{\includegraphics[width=0.24\textwidth]{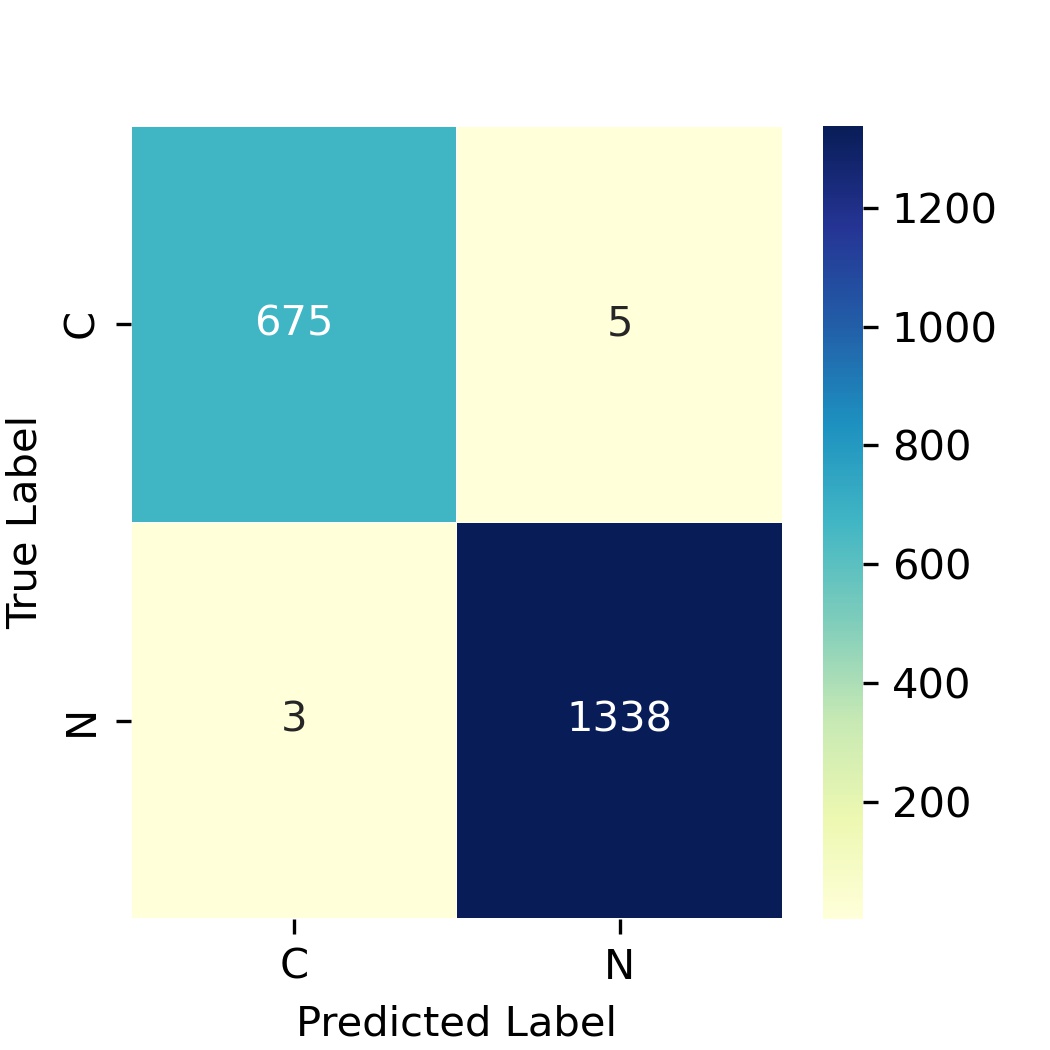}}
    \caption{Performance of proposed models on the Conv-Pneum dataset. The confusion matrix represents the detection rate of COVID-19 (C), Pneumonia (P), and Normal (N) (3-Class) in (a-d). (a) multilayer fusion of ResNet50V2 model, (b) multilayer fusion of InceptionV3 model, (c) singlelayer multimodal fusion model, and (d) multilayer multimodal fusion model. The detection rate of COVID-19 and Normal (2-Class) in (e-h). (e) multilayer fusion of ResNet50V2 model, (f) multilayer fusion of InceptionV3 model, (g) singlelayer multimodal fusion model, and (h) multilayer multimodal fusion model.}
    \label{Confusion-Matrix}
\end{figure*}

\subsection{Quantitative analysis}

The comparison of the proposed model with various models in terms of class-2 (COVID-19 vs Normal) and class-3 (COVID-19 vs Pneumonia vs Normal) classification accuracy is given in Table~{\ref{table:result}}. The proposed model outperformed the state-of-the-art algorithms and other sub-models by achieving class-2 and class-3 classification accuracy of 99.6\% and 97.21\%, respectively, and outperformed all existing models (p $<$ 0.05, Wilcoxon signed-rank test). Some enhancement is also achieved by the sub-model (proposed model without layer-wise fusion) compared to the second-best model Shanjiang et al.~{\cite{Trans-edl}}, where the sub-model achieved class-3 accuracy of 95.95\% in comparison with an accuracy of 95.69\%.

The evaluation of system performance always depends on each class classification result. Majorly, it is observed that some class classification results may be far away from the overall classification results, which is why we measure precision, recall, and F1 scores for each class of the proposed model. The F1-score, which is the harmonic mean of precision and recall, achieved by the proposed model for the class of COVID-19 is 0.94, for Pneumonia is 0.92, and for Normal is 0.98. Furthermore, the model obtained precision and recall values of 0.96, 0.94, 0.99, and 0.98, 0.96, 0.97, respectively for the classes of COVID-19, Pneumonia, and Normal. Table~{\ref{table:result}} shows that the proposed model surpassed the other state-of-the-art models and sub-models across all evaluation metrics. Therefore, it is evident that the system can accurately detect any class.

The confusion matrices of all four proposed models and sub-models in two categories of class-2 and class-3 are shown in Fig.~{\ref{Confusion-Matrix}}. In each matrix, the rows represent the actual class, and the columns represent the class predicted by each model. It was analyzed on the Conv-Pneum dataset, including 680 COVID-19, 1341 Normal, and 3875 Pneumonia CXR test images. The proposed multilayer multimodal model fusion model correctly identified 630 COVID-19 cases, and only 25 cases were misclassified out of 3875 Pneumonia cases, while in another instance, the singlelayer fusion model misclassified 257 cases. It indicates that the multilayer fusion model's error ratio is minimal compared to the other sub-models.

The performance of the proposed model is also evaluated on two other state-of-the-art datasets for validation purposes. The first dataset is the Dark-CovidNet~{\cite{DarkCovidNet}} dataset, on which the proposed model achieved a class-2 and class-3 classification accuracy of 98.25\% and 95.96\%, respectively, compared to Dark-CovidNet~{\cite{DarkCovidNet}} accuracy of 98.08\% and 87.02\%. The second dataset is the Rahimzadeh et al.~{\cite{RAHIMZADEH2020100360}} dataset, on which the proposed model achieved a class-2 accuracy of 98.25\% and class-3 accuracy of 95.96\%, respectively.

As shown in Table~{\ref{table:result}}, the proposed models outperformed the existing state-of-the-art models. Furthermore, we compared the accuracy reported by state-of-the-art works and found that our achieved accuracies are significantly higher than previous works. The comparison between reported accuracy and accuracy on the Cov-Pneum dataset of state-of-art models shows the changes. For instance, CoroNet~{\cite{CoroNet}} accuracy enhanced by around 3\% on our dataset, whereas CovidNet~{\cite{Covidnet}} detection accuracy degraded by about 1\%. Furthermore, Shanjiang et al.~{\cite{Trans-edl}} performance is increased in terms of accuracy (95.69\%) on the Cov-Pneum dataset compared to the reported accuracy of 94.53\%. However, our proposed model surpassed all the other models by achieving an accuracy of 99.60\% for class-2 classification and 97.21\% for class-3 classification, which presents our model as the most suitable classifier for CXR for the detection of COVID-19 and Pneumonia diseases.

\subsection{Qualitative analysis}

\begin{figure}
  \centering
   \includegraphics[width=0.48\textwidth,keepaspectratio]{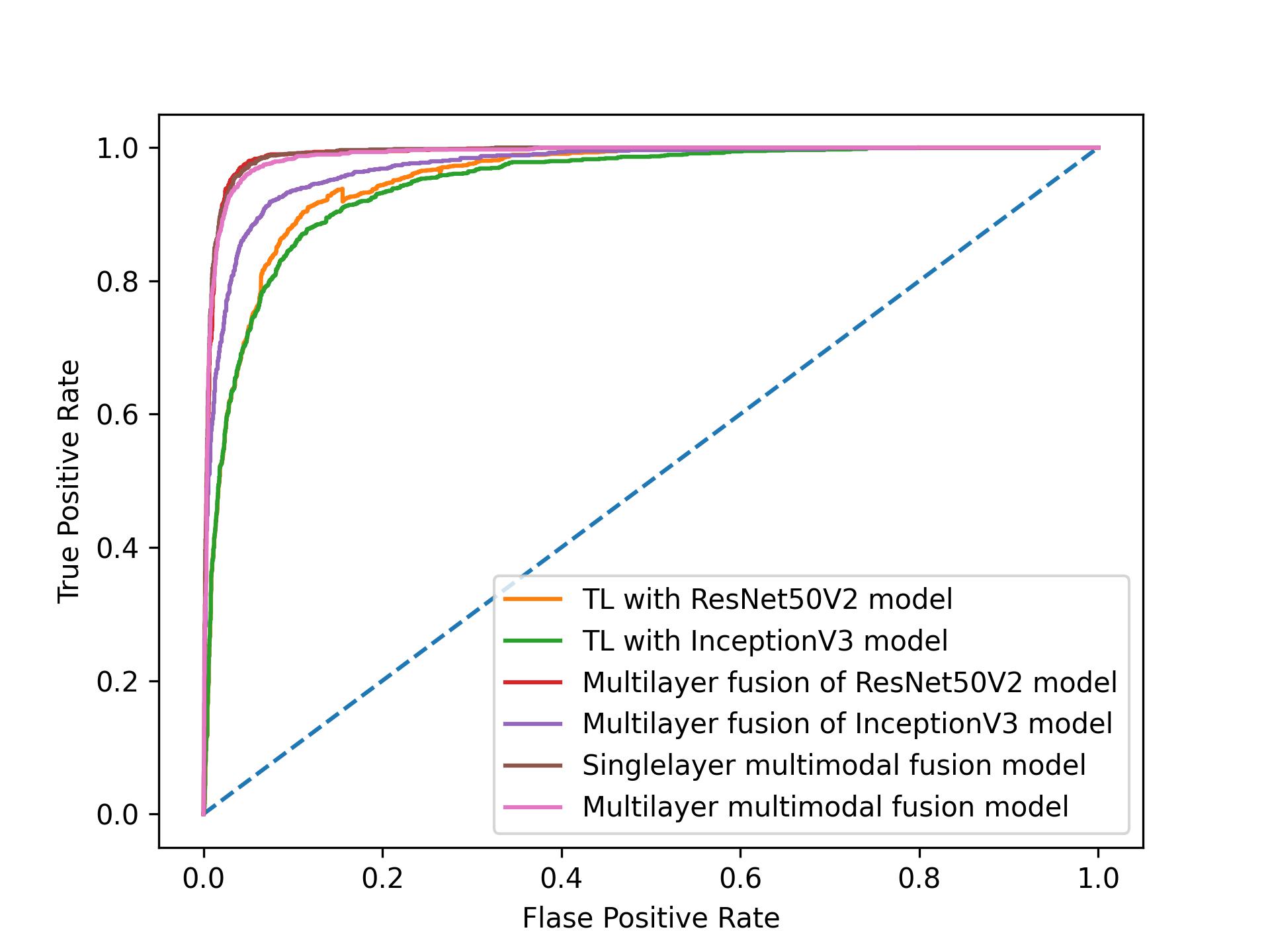}
  \caption{The ROC curve demonstrates the comparison between baseline models (TL with ResNet50V2~\cite{ResNet}, TL with InceptionV3~\cite{inception}) and proposed models in terms of AUC rate.}
  \label{ROC-Curve}
\end{figure}

\begin{figure}
  \centering
   \includegraphics[width=0.48\textwidth,keepaspectratio]{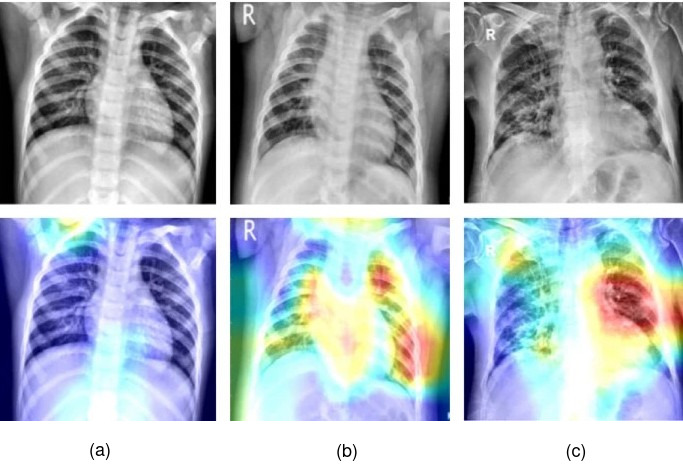}
  \caption{Grad-CAM visualization for CXR images: Original images (1st column) and overlaid activation maps (2nd column) for (a) Normal, (b) Pneumonia, and (c) COVID-19 cases.}
  \label{gradcam}
\end{figure}

The performance of the proposed model is measured through a receiver operating characteristic (ROC) curve, as shown in Fig.~{\ref{ROC-Curve}}. It demonstrates the comparison of True Positive Rate (TPR) vs False Positive Rate (FPR) between the proposed model and the baseline and sub-models. Generally, the larger area covered by the curve represents the highest accuracy model. It is visible that the proposed multilayer multimodal fusion model has the highest area compared to other comparable models. Moreover, to enhance the interpretability our proposed model, we utilized Gradient-weighted Class Activation Mapping (Grad-CAM), which effectively elucidates the model's decision-making process. This class activation map is generated by channeling gradient information from CXR images back into the final convolutional layer, allowing us to discern the significance of each neuron in classifying images into different disease classes.  Fig.~\ref{gradcam} showcases the class activation maps for CXR images of COVID-19, Pneumonia, and Normal cases. In these maps, light colors (such as blue) represent clear lungs, while hot colors (like red and yellow) indicate areas of infection in the lungs. Therefore, it is evident from the class activation maps of COVID-19 and Pneumonia that infected lungs display a hot color, while normal lungs appear in a lighter color (blue). These activation maps serve as a valuable tool for healthcare professionals to visualize the lungs and highlighting the infected areas. These results indicate that the proposed model is a better classifier for COVID-19 and Pneumonia diseases.

\begin{table}[]
\caption{Computational performance evaluation of state-of-art and proposed model.}
\label{tab:comp-table}
\begin{tabular}{@{}ll@{}}
\toprule
Methods & \begin{tabular}[c]{@{}l@{}}\#Param\\  (Million)\end{tabular} \\ \midrule
DarkCovid-Net~\cite{DarkCovidNet}  &  11.64 \\
 CoroNet~\cite{CoroNet} & 33.96  \\
CovidNET~\cite{Covidnet}  & 11.75  \\
CovidGAN~\cite{Covidgan} & 14  \\
ResNet50V2 & 24.97  \\
InceptionV3 & 23.8  \\
\textbf{Multilayer multimodal fusion model} & \textbf{16.01 } \\ \bottomrule
\end{tabular}
\end{table}

\section{Discussion}\label{7}

This study introduces a novel fusion model that leverages the strengths of the ResNet50V2 and InceptionV3 architectures, resulting in enhanced effectiveness and improved classification reliability and accuracy. The main contribution of this research lies in exploring various fusion scenarios, including singlemodal vs multimodal and singlelayer vs multilayer, with significant performance gains observed in the multilayer multimodal fusion approach. Additionally, the proposed methodology successfully addresses the challenge of fusing feature maps of different sizes, improving the model's adaptability to diverse data.

One significant accomplishment of this research is the creation of the Cov-Pneum dataset, which surpasses the size of existing state-of-the-art datasets by more than ($>$16-95\%). This larger dataset mitigates data sparsity issues and strengthens the robustness of our findings.

In the evaluation of the proposed model's performance with varying dropout values (0.2, 0.25, 0.3, 0.35, and 0.4), it was observed that the model achieved its highest class-3 accuracy of 97.21\% when a dropout rate of 0.3 was applied. This result suggests that a dropout rate of 0.3 struck a balance between preventing overfitting and preserving valuable features in the network, leading to the most optimal performance. Lower dropout rates, such as 0.2 and 0.25, were associated with increased overfitting, resulting in lower accuracy of 94.2\% and 95.58\%, respectively, while higher dropout rates of 0.35 and 0.4 overly regularized the model, causing a drop in accuracy due to excessive feature dropout, resulting in lower accuracy of 96.25\% and 94.15\%, respectively. Therefore, the optimized value of dropout rate is 0.3 taken in this study leading to the best performance in lung disease classifications.

In-depth experiments are conducted to evaluate the performance of the fusion approach, and it is observed that the four-layer fusion yields outstanding results compared to singlelayer or three-layer fusion. However, fusing the fifth layer leads to degraded results due to redundant feature maps and increased computational overhead. Hence, we adopt the four-layer fusion for our experiments.

An essential advantage of the proposed model is its computational efficiency, outperforming existing methods in the number of trainable parameters. It is evident from Table~{\ref{tab:comp-table}} that the number of trainable parameters used in the proposed model is the least as compared to base models (ResNet and Inception). We have successfully preserved the computational complexity after integrating the two deep models. The proposed model has 16.01 million parameters, approximately 67\%  of the parameters used in the computationally best available base model. This substantial reduction allows our architecture to balance computational efficiency and performance, making it an attractive solution for resource-constrained scenarios where fast and efficient processing is critical.

While deep learning-based models have been used for detecting COVID-19 and Pneumonia lung diseases, our proposed model extends its applicability to the healthcare industry. It can be deployed as an automated disease classifier at the primary stage, allowing doctors to save time initiating treatment. Moreover, it shows potential for use in other disease classification tasks and may serve as an alternative in areas with limited access to radiologists.

\subsection{Future directions of the proposed research}

The quantitative and qualitative analysis suggests that our model is robust and has the potential to be deployed in the healthcare system for accurate disease classification and reducing the workload of radiologists. To further extend the applicability of this research in the medical field, future directions can explore the generalization capability of the model on other lung diseases, such as Lung Cancer, Tuberculosis, and Pneumothorax. Moreover, incorporating multimodal data, such as combining blood parameters with radiological imaging, holds promise in improving detection accuracy and providing a more comprehensive analysis of lung diseases. Exploring these possibilities can lead to advancements in disease diagnosis and management, ultimately benefiting patient care and healthcare systems.

\section{Conclusions}\label{8}

In this study, we present a novel CNN-based fusion model tailored for the classification of chest X-ray images. The primary goal of our research was to enhance the accuracy and reliability of automated classification systems used in diagnosing lung diseases, including COVID-19, Pneumonia, and Normal cases. The proposed approach involves the multilayer fusion of ResNet50V2 and InceptionV3 models, combined with multimodal fusion. Unlike previous studies that focused on individual models or fusion techniques, our research emphasizes the synergistic effects achieved through multilayer multimodal fusion using state-of-the-art CNN architectures. Additionally, we have introduced an independent feature map transform module to effectively fuse feature maps of different sizes, resolving the variable-sized problem encountered in generating multiple layers of feature maps. The experimental results have demonstrated the promising potential of our approach in advancing CXR classification, yielding improved accuracy and reliability. This work opens new  possibilities for computer-aided diagnosis (CAD) systems, revolutionizing the field of medical image analysis.

\section*{Compliance with Ethical Standards}
The work is original and not submitted elsewhere. The authors do not have any conflict of interest. This article does not contain any studies with human participants or animals performed by any of the authors.

\section*{Data availability statement}

The original contributions presented in this study are publicly available. This data and source code are available on https://github.com/MIILab-IITGN/MultiFusionNet.

 \section*{Acknowledgment}
YKM was supported by Indian Institute of Technology Gandhinagar startup grant IP/IITGN/CSE/YM/2324/05. SA and KVA are supported by the ABV-Indian Institute of Information Technology and Management, Gwalior, India 

\section*{Author contributions}
SA, KVA, and YKM conceptualized and designed the methodology. SA prepared the dataset and conducted the analyses. YKM and KVA supervised the study. SA and YKM validated the results. YKM wrote the first draft of the manuscript. All authors interpreted the data and revised the manuscript.



\bibliography{sn-bibliography}


\end{document}